\documentclass[journal]{IEEEtai}
\usepackage[colorlinks,urlcolor=blue,linkcolor=blue,citecolor=blue]{hyperref}
\usepackage{color,array}

\usepackage{amsmath,amssymb,amsfonts}
\usepackage{algorithmic}
\usepackage{graphicx}
\usepackage{textcomp}

\usepackage{makecell} 
\usepackage{multirow} 

\usepackage[backend=bibtex,style=ieee,natbib=true,sorting=none,urldate=comp]{biblatex} 
\usepackage{tikz} 
\usetikzlibrary{calc}

\begin{document}

\title{Deep Reinforcement Learning and Convex Mean-Variance Optimisation for Portfolio Management}

\author{Ruan Pretorius, Terence L. van Zyl, \IEEEmembership{Member, IEEE}
\thanks{The support of the DSI-NICIS National e-Science Postgraduate Teaching and Training Platform (NEPTTP) towards this research is hereby acknowledged. Opinions expressed and conclusions arrived at, are those of the authors and are not necessarily to be attributed to the NEPTTP.}
\thanks{Ruan Pretorius is with the School of Computer Science and Applied Mathematics, University of the Witwatersrand, South Africa and the DSI-NICIS National e-Science Postgraduate Teaching and Training Platform (NEPTTP), South Africa (e-mail: pretoriusr.7@gmail.com).}
\thanks{Terence L. van Zyl is with the Institute for Intelligent Systems, Unversity of Johannesburg, Johannesburg, South Africa (e-mail: tvanzyl@gmail.com).}
\thanks{This paragraph will include the Associate Editor who handled your paper.}}

\markboth{Journal of IEEE Transactions on Artificial Intelligence, Vol. 00, No. 0, Month 2021}
{Ruan Pretorius \MakeLowercase{\textit{et al.}}: Deep Reinforcement Learning and Convex Mean-Variance Optimisation for Portfolio Management}

\maketitle

\begin{abstract}
Traditional portfolio management methods can incorporate specific investor preferences but rely on accurate forecasts of asset returns and covariances. Reinforcement learning (RL) methods do not rely on these explicit forecasts and are better suited for multi-stage decision processes. To address limitations of the evaluated research, experiments were conducted on three markets in different economies with different overall trends. By incorporating specific investor preferences into our RL models' reward functions, a more comprehensive comparison could be made to traditional methods in risk-return space. Transaction costs were also modelled more realistically by including non-linear changes introduced by market volatility and trading volume. The results of this study suggest that there can be an advantage to using RL methods compared to traditional convex mean-variance optimisation methods under certain market conditions. Our RL models could significantly outperform traditional single-period optimisation (SPO) and multi-period optimisation (MPO) models in upward trending markets, but only up to specific risk limits. In sideways trending markets, the  performance of SPO and MPO models can be closely matched by our RL models for the majority of the excess risk range tested. The specific market conditions under which these models could outperform each other highlight the importance of a more comprehensive comparison of Pareto optimal frontiers in risk-return space. These frontiers give investors a more granular view of which models might provide better performance for their specific risk tolerance or return targets.
\end{abstract}

\begin{IEEEImpStatement}
Our RL models address limitations found in previous research by including non-linear terms in the transaction cost model, introducing investor preference parameters in the models' reward functions, and repeating experiments on different market conditions. These contributions improve the accuracy of experimental simulation, allow for a more comprehensive comparison to traditional mean-variance optimisation methods, and provide a tool for investors that takes their specific risk tolerance into account. Depending on the overall price trend of the market, our RL models were either able to match or significantly outperform traditional mean-variance models for the majority of risk values tested.
\end{IEEEImpStatement}


\section{Introduction}
\label{sec:introduction}


\PARstart{P}{ortfolio management} (also called portfolio optimisation or asset allocation) is the process of allocating portions of some total amount of wealth to different assets in an asset universe. Modern portfolio theory and the concept of portfolio management was first introduced by Harry Markowitz in the 1950s~\cite{x,y}. The aim of portfolio management is to optimally assign a proportion of wealth to each asset in an asset universe so that some goal is achieved (usually to maximise expected returns or limit risk over some investment period). There are typically more than one time-step considered within an investment period where the allocation of assets can be adjusted or rebalanced as more recent information becomes available. This rebalancing is done in order to keep the portfolio performance in line with the investor's preferences~\cite{o}.

One of the main reasons an investor might want to distribute their wealth between a variety of assets in a portfolio, as opposed to investing only in a single asset, is that it promotes diversification and mitigates risk~\cite{x,y}. Zivot (2017) has proven that the volatility (risk) of a long-only portfolio of assets is always lower than that of a single asset, given the assets in the portfolio are not perfectly correlated~\cite{z}.


Harry Markowitz's framework of \textit{mean-variance} portfolio optimisation is widely used in industry and academia. It allows an investor to optimally allocate their wealth between assets in order to balance the risk-reward trade-off according to their risk appetite. For example, an investor might decide on a maximum amount of risk that they are willing to tolerate. Markowitz's mean-variance method allows them to choose the optimal weighting of assets in a portfolio to maximise their expected returns without that level of risk being exceeded. When these optimal portfolios are computed for a range of different risk values and plotted in risk-return space, they form curve called the \textit{efficient frontier}. This efficient frontier can be used to select the optimal portfolio with maximum expected returns for a given risk value~\cite{y,z}.

One of the main limitations of Markowitz's mean-variance method is that it only considers one time-step into the future~\cite{o}. In other words, the allocation of assets is done in a way that only takes a single portfolio rebalancing period into account. Ideally, the impact on future decisions should also be taken into account.

Another limitation of traditional mean-variance methods is that they rely on accurate forecasts of returns and covariances between the assets that make up the portfolio. Unrealistic assumptions of normally distributed returns are also sometimes made which can lead to large drawdowns that often cannot be tolerated by some investors~\cite{g}.

In 2017, a study by Boyd \textit{et al.} showed how this single-period optimisation (SPO) approach can be extended to a multi-period optimisation (MPO) version~\cite{boyd}. Both the SPO and MPO versions of Boyd \textit{et al.} (2017) are used as traditional mean-variance benchmark methods in this study. These methods are convex optimisation problems that aim to maximise expected returns in the presence of transaction costs and risk.

\subsection{Background}
\subsubsection{Reinforcement Learning}
\begin{figure}[th]
    \centerline{
    \begin{tikzpicture}
      \draw (0,0) node(env) [rectangle, draw=black, thick, minimum height=8mm, minimum width=25mm] {Environment};  
      \draw (0,2) node(agent) [rectangle, draw=black, thick, minimum height=8mm, minimum width=25mm] {Agent};
      \draw[-stealth,thick] (agent.east) -| (3,1) node(a)[anchor=east]{action} |- (env.east); 
      \draw[-stealth,thick] ($(env.west)+(0mm,2mm)$) -| ($(-3,1)+(2mm,0mm)$) node(r)[anchor=west]{reward} |- ($(agent.west)+(0mm,-2mm)$); 
      \draw[-stealth,thick] ($(env.west)+(0mm,-2mm)$) -| ($(-3,1)+(-2mm,0mm)$) node(s)[anchor=east]{state} |- ($(agent.west)+(0mm,2mm)$); 
    \end{tikzpicture}}
    \caption{Interaction between agent and environment formulated as a Markov decision process.}
    \label{fig:rl-mdp}
\end{figure}
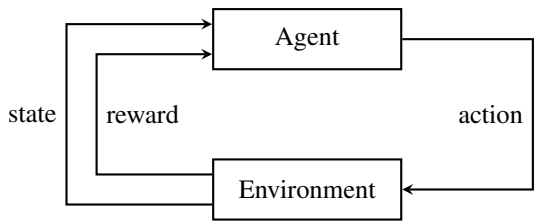

Sutton and Bartow (2018) gave the following description of reinforcement learning (RL)~\cite{sutton-bartow}. It describes both a type of problem and a class of solutions that work well to solve that problem. It applies to problems and solutions that can be formulated in terms of an agent that interacts with and receives feedback from its environment. This interaction is often framed as a Markov decision process (MDP) which has three essential components. These are sensations in the form of observable states of the environment, actions that can be executed by the agent, and reward signals that guide the agent towards a goal related to an ideal state of the environment. The agent is never given explicit instructions of which actions to take but instead learns which actions are best to take in different circumstances through exploration of its environment. Fig.~\ref{fig:rl-mdp} shows a visual representation of this agent-environment interaction.

By formulating sequential decision processes in terms of MDPs, future states, actions, and rewards depend on past ones. Therefore, the MDP formulation captures the need for a trade-off between immediate and delayed rewards. RL methods are able to solve these types of problems  because the goal of the agent is to maximise expected future rewards~\cite{sutton-bartow}.

\subsubsection{Reinforcement Learning for Portfolio Management}
To emphasise the applicability of RL methods to the portfolio management task, the relevant MDP components are identified here. The role of the agent is that of a decision-maker that has to choose the optimal weighting between assets in a portfolio (action). The environment in this case is the market, which supplies the observable states (such as historical prices) and reward signals (such as realised returns after transaction costs) as feedback.

When transaction costs are considered, the portfolio management problem becomes a multi-stage decision-making process where future states and decisions are impacted by past decisions~\cite{n}. In this setting, immediate rewards are not the only important aspect to consider, but also the possible negative impact of current decisions on the ability to receive rewards in the future. RL methods are particularly well suited for this type of problem since they aim to maximise the accumulation of rewards in the long term even if it means acting sub-optimally in the short term~\cite{sutton-bartow}. This capacity to make long-term decisions is the main reason for choosing RL methods as the subject of investigation in the portfolio management task.

\subsubsection{Related Research}
Jiang \textit{et al.} (2017) conducted a study using deep RL for portfolio management on a 12-asset cryptocurrency portfolio~\cite{j}. They used three different deep neural network architectures for function approximation. Their best performing model contained a convolutional neural network (CNN) which produced a Sharpe ratio of 0.087, a maximum drawdown of 22\%, and 400\% returns in a 50-day period  while considering transaction costs with a commission rate of 0.25\%~\cite{j}.

Another study, conducted by Filos (2019)~\cite{meng}, used different types of RL methods on a 12-asset portfolio consisting of cash and stocks from the S\&P500 universe. They devised a new method called Deep Soft Recurrent Q-network (DSRQN) which was an adaption of both the Deep Q-network (DQN) of Mnih \textit{et al.} (2015)~\cite{aa} and the deep recurrent Q-network (DRQN) of Hausknecht and Stone (2015)~\cite{bb}. This DSRQN method could produce 256\% returns with a Sharpe ratio of 2.4 and a maximum drawdown of 85\% over the 13-year test period. Another method used in the same study by Filos (2019), was a Monte Carlo policy gradient method called REINFORCE. It could produce returns of 325\% with a Sharpe ratio of 3.02 and a maximum drawdown of 63.5\% over the 13-year test period~\cite{meng}. All methods in the study by Filos (2019) considered transaction costs in the form of broker commission and spreads lumped together as 0.2\% of the change in portfolio weights~\cite{meng}.

A more recent study by Yang \textit{et al.} (2020) looked at an ensemble of three different deep reinforcement actor-critic methods on a portfolio of the 30 stocks of the Dow Jones Industrial Average (DJIA) index. This ensemble method could produce 13\% annual returns with an annual volatility of 9.7\%,  an annual Sharpe ratio of 1.3, and a maximum drawdown of 9.7\% over a period of four and a half years~\cite{p}.

\subsection{Current Limitations}
The preceding literature demonstrates that researchers have previously investigated numerous approaches to portfolio management with RL. This usage includes different combinations of on-policy and off-policy learning with temporal difference (TD) methods and Monte Carlo methods for value-function estimation, policy estimation, and actor-critic methods. However, this paper identifies the following four limitations in the evaluated research.

Firstly, most of the RL methods were not risk-aware due to the reward signal being related exclusively to returns (except for those methods that used the Sharpe ratio as a reward function). The large drawdowns during test periods confirm the risk-ignorant nature of some of these models. Most of the evaluated research aimed to outperform the market or traditional methods only in terms of returns. This maximum return aim does not cater to the needs of different investors with different risk tolerances and return targets. For example, more risk-averse investors that are not necessarily aiming for maximising returns irrespective of risk might want to incorporate some limit to the risk they are willing to assume by investing in these risky assets.

The second and closely related limitation of the evaluated research was that they only compared single portfolio outcomes. Apart from only being of interest to investors with particular risk and return goals, this gives a limited view of the model's performance in the risk-return space. In other words, the evaluated studies produced only a single performance point in the risk-return space instead of an entire efficient frontier.

Thirdly, in the aforementioned research, the transaction cost was limited in that it was a linear function of bid-ask spreads and broker commission. This limitation neglects the inclusion of a second, non-linear term that changes as a function of market volatility and trading volume, which is a more realistic characterisation of transaction costs~\cite{boyd}.

Finally, the above research only assessed the performance of models on a single market. Therefore, they were limited in that their results might not apply to markets in different economies or markets with different characteristics and conditions as far as overall market behaviour is concerned.


\subsection{Contribution}
This study compared the portfolio management performance of RL methods to traditional mean-variance methods while addressing the limitations mentioned above present in previous research. To this end, non-linear terms were included in the transaction cost model to improve the accuracy of experimental simulation of RL models. Different investor preference parameters were also included in the RL models' reward functions. The inclusion of these parameters enabled a more comprehensive comparison between RL models and traditional mean-variance optimisation models in the risk-return space.  All experiments in this study were repeated three times using data from different markets to give a more granular insight into the different market conditions which can influence the dominance of one method over others.

\section{Method}
\subsection{Portfolio Performance Measures}
Because a portfolio is composed of a collection of assets whose prices change over time, the portfolio's value also changes over time. The returns and risk exposure changes can be quantified and depend on the prices of the underlying assets in the portfolio and the weighting assigned to each one. In order to understand and appreciate the methods and results of this study, this section presents an overview of portfolio performance measures to provide additional context.

The portfolio performance measures used in this study were returns, volatility, and Sharpe ratio. These measures are commonly found in literature and are described in this paper using notation similar to that of Boyd \textit{et al.} (2017)~\cite{boyd}.

For a portfolio of $n$ assets (stocks) and cash, the price vector at any time-step $t$ is denoted $p_t \in \mathbb{R}^{n+1}_{+}$. From this price vector, a vector of returns $r_t \in \mathbb{R}^{n+1}$ can be constructed. The return of asset $i$ is the percentage price change between two successive time-steps: 
\begin{equation}
\label{eq:ret-asset}
    (r_t)_i = \frac{(p_t)_i - (p_{t-1})_i}{(p_{t-1})_i}, \quad i=1,\dots,n+1
\end{equation}

Alternatively, the log-return is also sometimes used and is calculated as follows:
\begin{equation}
\label{eq:log-ret}
    \log \left( \frac{(p_t)_i}{(p_{t-1})_i} \right) = \log \left( 1+(r_t)_i \right), \quad i=1,\dots,n+1
\end{equation}

At any time-step $t$, the proportion of the total portfolio assigned to each asset is captured by the portfolio weight vector $w_t \in \mathbb{R}^{n+1}$. It is useful to also define the change in weight $z_t \in \mathbb{R}^{n+1}$ between successive time-steps for each asset $i$ as follows:
\begin{equation}
\label{eq:weight-change}
    (z_t)_i = (w_t)_i - (w_{t-1})_i, \quad i=1,\dots,n+1
\end{equation}

In this study, as in Boyd \textit{\textit{et al.}} (2017)~\cite{boyd}, transaction costs were modelled as a unit-less, non-linear function of bid-ask spread, broker commissions, trading volume and market volatility. The total transaction cost $\phi^{\text{trade}}_{t}$ at time $t$ is the sum of transaction costs resulting from trading individual assets as described in Equation~\eqref{eq:transaction-cost}. Note that a double subscript is used to indicate time-step and asset (e.g. $z_{t,i} = (z_t)_i$).
\begin{equation}
\label{eq:transaction-cost}
    \phi^{\text{trade}}_{t} = \sum_{i=1}^{n} \left[ a |z_{t,i}| + b \sigma_{t,i} \frac{|z_{t,i}|^{3/2}}{\sqrt{V_{t,i}/v_t}} +  c z_{t,i} \right]
\end{equation}

Here,  $a \in \mathbb{R}$ is used to capture half of the bid-ask spread expressed as a fraction of asset price. Any broker commission can also be incorporated in $a$. Here, $b \in \mathbb{R}$ has units of inverse-dollars and is used to scale the second term. The recent volatility (standard deviation of returns) of asset $i$ at time $t$ is captured by $\sigma_{t,i} \in \mathbb{R}$ in dollars. $V_{t,i} \in \mathbb{R}$ is the volume traded in dollars of asset $i$ at time $t$, which is scaled by the total portfolio value $v_t$ in order to keep the denominator unit-less. Finally, $c \in \mathbb{R}$ can be used to create asymmetry in the transaction cost when buying and selling do not cost the same.

The realised return $R_t^p$ (after transaction costs) of the entire portfolio of $n+1$ assets at time $t$ can then be calculated as follows:
\begin{equation}
\label{eq:ret-portfolio}
    R_t^p = r_t^Tw_t + r_t^Tz_t - \phi^{\text{trade}}_{t}
\end{equation}

When considering some investment period from ${t=1,\dots,T}$, it is common to calculate the average realised returns $\overline{R^p}$ for that period as follows:
\begin{equation}
\label{eq:avg-real-rets}
    \overline{R^p} = \frac{1}{T} \sum_{t=1}^{T} R_t^p
\end{equation}

The risk associated with holding a portfolio can be quantified by the standard deviation of portfolio returns $\sigma^p$. This value is commonly referred to as volatility and can be calculated using the following equation:
\begin{equation}
\label{vol-portfolio}
    \sigma^p = \left[ \frac{1}{T} \sum_{t=1}^{T} \left( R_t^p - \overline{R^p} \right)^2 \right]^{1/2}
\end{equation}

It is often useful to compare the portfolio risk and return to some benchmark. This benchmark can either be another portfolio or a single asset. The terms \textit{excess return} and \textit{excess risk} are used to refer to the risk and return obtained in excess of a risk-free asset like cash. Excess return $R_t^e$ and of a portfolio is defined in Equation~\eqref{ret-excess} and the excess risk $\sigma^e$ can be calculated as the standard deviation of excess returns.
\begin{equation}
\label{ret-excess}
    R_t^e = R_t^p - (r_t)_{n+1}
\end{equation}

Here, $(r_t)_{n+1}$ refers to the return of the $(n+1)^{\text{th}}$ asset of the portfolio (the risk-free or cash asset).

The Sharpe ratio $SR$ is used to quantify the risk-adjusted excess returns of a portfolio, described as follows:
\begin{equation}
\label{sharpe}
    SR = \frac{\overline{R^e}}{\sigma^e}
\end{equation}

With these portfolio performance metrics established, some traditional methods of portfolio management can be considered.

\subsection{Traditional Mean-Variance Optimisation Baselines}

As mentioned in the introduction, SPO and MPO extensions of Markowitz's mean-variance optimisation, developed by Boyd \textit{et al.} (2017)~\cite{boyd}, were used as baseline portfolio management methods in this study. These methods are convex optimisation problems formulated to enable single and multi-period optimisation. 

The SPO version arrives at the optimal portfolio weight vector for the next time-step $w_{t+1} = w_t + z_t$ by solving for $z_t$ in the optimisation problem:
\begin{align}
\label{eq:spo}
    \text{max} \quad &\hat{r}^T_t (w_t + z_t) - \gamma^{\text{trade}} \hat{\phi}^{\text{trade}}_{t} (z_t) \nonumber \\
    &- \gamma^{\text{risk}} \psi_t (w_t + z_t) \\
    \text{s.t.} \quad &z_t \in \mathcal{Z}_t, \quad w_t + z_t \in \mathcal{W}_t, \quad \mathbf{1}^T z_t = 0 \nonumber
\end{align}
where $\psi_t (w_t + z_t)$ describes the risk function which which is an estimate of the variance of portfolio returns. A caret is placed over some variables to emphasise that they are estimates (since they are not known at time $t$). Here, $\gamma^{\text{trade}}$ and $\gamma^{\text{risk}}$ scale the trading and risk aversion respectively and can be changed to capture the preferences of any investor. As the trading aversion increases, trading will be discouraged and transaction costs will decrease. The risk aversion parameter is used to discourage holding portfolios with high volatility. In this study the quadratic risk function is used and is described as follows:
\begin{equation}
\label{eq:risk-function}
    \psi_t(w_t + z_t) = (w_t + z_t)^T \hat{\Sigma}_t (w_t + z_t)
\end{equation}
where $\hat{\Sigma}_t \in \mathbb{R}^{(n+1)\times(n+1)}$ is an estimate of the return covariance matrix of all assets in the portfolio.

The constraints in Equation~\eqref{eq:spo} are used, for example, to ensure that only long trades are executed and to ensure that the sum of portfolio weights add to unity.

Similarly, the objective of the MPO version is to choose the change in portfolio vector that maximises expected realised returns while minimising risk and transaction costs. However, the MPO version extends the SPO framework to take multiple time-steps into account. This MPO constitutes a trading plan for $H$ time-steps into the future, producing a sequence of portfolio vector changes, $z_t,z_{t+1},\dots,z_{t+H-1}$ by solving:
\begin{align}
\label{eq:mpo}
    \text{max} \quad &\sum_{\tau=t}^{t+H-1}  \Big[ \hat{r}^T_{\tau|t} (w_\tau + z_\tau) - \gamma^{\text{trade}}  \hat{\phi}^{\text{trade}}_{\tau} (z_\tau) \nonumber \\ 
    &\quad\quad\quad\quad - \gamma^{\text{risk}} \psi_\tau (w_\tau + z_\tau) \Big]  \\
    \text{s.t.} \quad &z_t \in \mathcal{Z}_t, \quad w_t + z_t \in \mathcal{W}_t, \quad \mathbf{1}^T z_t = 0, \nonumber \\
    & w_{\tau+1} = w_\tau + z_\tau, \quad \tau = t,\dots,t+H-1 \nonumber
\end{align}
where $\hat{r}_{\tau|t}$ is used to denote the return forecast of $r_\tau$ made at time $t$, using only information available at time $t$. In both the SPO and MPO versions, $w_t$ is known since it is the current portfolio weight vector.

Following the method of Boyd \textit{et al.} (2017), the MPO version used in this study was a two-period optimisation where $H=2$. The same values were also used for $\gamma^{\text{risk}}$ and $\gamma^{\text{trade}}$ (with some extension on both extreme ends) so that all 504 pairwise combinations of the following sets were tested to capture a wide range of investor preferences:
\begin{align*}
    \gamma^{\text{risk}} \in \{&0.1, 0.178, 0.316, 0.562, 1, 2, 3, 6, 10, 18, 32, 56, \\&100, 178, 316, 562, 1000, 2000, 5000, 10000, \\&20000\} \\
    \gamma^{\text{trade}} \in \{&0.1, 0.5, 1, 2, 3, 4, 5, 5.5, 6, 6.5, 7, 7.5, 8, 9, 10, 11, \\&12, 15, 20, 30, 45, 60, 100, 200\}
\end{align*}

As in Boyd \textit{et al.} (2017), the asset returns covariance matrix $\hat{\Sigma}$ was estimated using a factor model~\cite{boyd}. This involved, firstly, calculating the actual returns covariance matrix of the trailing two-year period $\Sigma^{\text{past}}$. Secondly, an eigendecomposition of this covariance matrix of past returns was performed as follows:
\begin{equation}
    \Sigma^{\text{past}} = \sum_{i=1}^n \lambda_i q_i q_i^T
\end{equation}
where the eigenvalues $\lambda_i$ were in descending order. Thirdly, these were used to construct the covariance matrix of factor returns ${\Sigma^f=\text{diag}(\lambda_1,\dots,\lambda_1)}$, the factor loading matrix ${F=\left[q_1,\dots,q_k \right]}$, and the idiosyncratic risk matrix ${D=\sum_{i=k+1}^n\lambda_{i}\text{diag}(q_i)\text{diag}(q_i)}$. Finally, these components were used to construct a factor model to estimate the asset returns covariance matrix as follows:
\begin{equation}
    \hat{\Sigma} = F \Sigma^f F^T + D
\end{equation}
where ${\hat{\Sigma} \in \mathbb{R}^{(n+1)\times(n+1)}}$, ${F \in \mathbb{R}^{(n+1)\times k}}$, ${\Sigma^f \in \mathbb{R}^{k \times k}}$, and ${D \in \mathbb{R}^{(n+1)\times(n+1)}}$. In this study, $k=15$ factors were used. Using this factor model enabled faster simulation times on the order of $\mathcal{O}(nk^2)$ as opposed to $\mathcal{O}(n^3)$ when not using a factor model~\cite{boyd}.

Another model reported on in this paper as a benchmark was the \textit{equally weighted} (EW) model. This model did not rely on any return forecasts or other estimations of the underlying assets in the portfolio but instead applied a single, simple rule to make allocation decisions. The EW model started off fully invested in non-cash assets, where each asset was assigned an equal weight, i.e., $w_t = \left[ 1/n, 1/n, \dots, 0 \right]$. At the end of each day, the EW model rebalanced its holdings to re-establish this equal weighting.

\subsection{State-of-the-art RL Models for Comparison}
The investigation used three existing state-of-the-art actor-critic RL models as a baseline comparison for our RL models. These were Advantage Actor-Critic (A2C), Proximal Policy Optimisation (PPO), and Deep Deterministic Policy Gradient (DDPG). These were the same three models implemented in the study by Yang \textit{et al.} (2020)~\cite{p} in their ensemble model. The only modification made to these models were to change the linear transaction cost function so that it captures non-linear dynamics and matches Equation~\eqref{eq:transaction-cost}. We made this modification to allow for a valid comparison between all the models in this study. Apart from this modification, we kept all other aspects of the original models. This replication of the research ensured that these models were as close to their original forms as possible.

The critic network in A2C approximates what is called an advantage function in addition to the usual value function. This advantage function enables A2C to assess both the quality of actions and how good they can be, which leads to a more robust policy with lower variance. The experiments of Yang \textit{et al.} (2020) suggest that A2C performs better in down-trending markets with high volatility compared to both PPO and DDPG~\cite{p}.

DDPG combines Q-learning and policy gradient methods and has a policy network that deterministically maps states to actions. DDPG has a replay buffer that stores the state transitions and their corresponding actions and rewards during training. Based on batches of these transitions drawn from the replay buffer, the model parameters are updated~\cite{p}.

Finally, PPO introduces a clipping term to its loss function that discourages large policy changes when model parameters are updated, leading to more stable policy learning. According to the experimental results of Yang \textit{et al.} (2020), both DDPG and PPO perform better in sideways and upward-trending market conditions compared to A2C, with PPO slightly outperforming DDPG~\cite{p}.

The continuous state space for DDPG, PPO, and A2C included technical indicators containing the historical price, trend, volume information, and the current portfolio holdings to allow them to account for transaction costs. The action space was also continuous for all three models, consisting of a normalised vector that specifies the number of shares to buy or sell for each asset in the portfolio. As in the original study, we set the reward function of these models to maximise realised portfolio returns~\cite{p}. We also set all hyperparameter values for these models to the default values used in the code library accompanying the original study~\cite{ff}.

\subsection{Our RL Models}
Our RL model was named FRONTIER (rein\textbf{F}orcement lea\textbf{R}ning p\textbf{O}rtfolio ma\textbf{N}ager wi\textbf{T}h \textbf{I}nv\textbf{E}stor p\textbf{R}eferences) due to its capacity to take different investor preferences into account and its output creating a Pareto optimal frontier in risk-return space (explained below).
FRONTIER is a Monte Carlo policy gradient method based on the REINFORCE algorithm~\cite{sutton-bartow}. The policy of this model was represented with a neural network (policy network). Three different policy networks were examined in this study in order to determine the limitations and advantages of each one. For all policy networks, the observable state input variables or features were supplied as an input at the beginning of each time step. Although different intermediate layers were used for the different policy networks, they all had an output layer with a softmax activation function and one neuron for each asset in the portfolio. This final output layer thus produced the portfolio weight vector $w_{t+1}$ to be used in the next time step. Therefore, FRONTIER had a continuous action space in $\mathbb{R}^{n+1}_{+}$ (the subscript indicates non-negative values, which implies long-only policies).

\begin{figure*}[th]
\centerline{\includegraphics[width=0.65\textwidth]{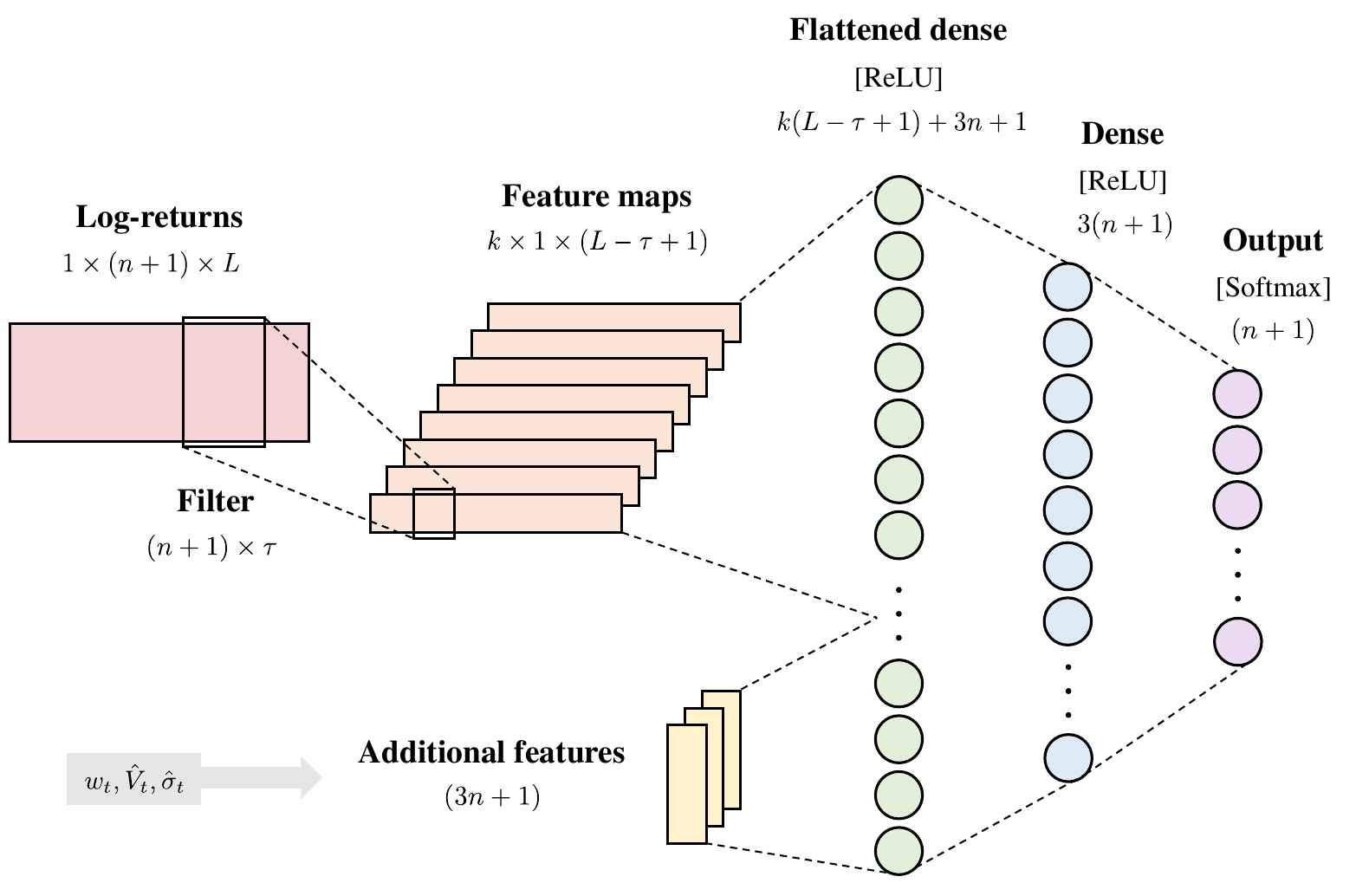}}
\caption{The log-returns policy network of the FRONTIER model with only historical log-returns and additional features as state inputs. A convolutional filter was passed over the log-returns window to produce several feature maps. The additional feature vectors gave the model the capacity to take transaction costs into account. The feature maps and additional features were flattened and connected with fully-connected layers to produce the next portfolio weight vector as an output.}
\label{fig:policy-cnn}
\end{figure*}

The first variation of the policy network, seen in Fig.~\ref{fig:policy-cnn}, was called the \textit{log-returns} policy.

For this policy network, a window of historical log-returns was passed as input along with some additional features. The log-returns window was made up of the daily log-returns for each asset in the portfolio, spanning a length of $L$ time-steps, calculated using Equation~\eqref{eq:log-ret}. A convolutional filter of size $(n+1) \times \tau$ was then passed over this log-returns window to produce $k$ feature maps. This convolutional neural network (CNN) block automatically detects patterns in individual assets and between assets (such as covariance). 

In addition to the log-returns, three additional feature vectors were also given as state inputs to give the model the capacity to take transaction costs into account (note these features correspond to factors and terms in the transaction cost Equation~\eqref{eq:transaction-cost}). These three features were the current portfolio weight vector $w_t \in \mathbb{R}^{n+1}_{+}$, the estimated volume traded for each non-cash asset $\hat{V}_t \in \mathbb{R}^{n}_{+}$, and the estimated volatility for each non-cash asset $\hat{\sigma}_t \in \mathbb{R}^{n}_{+}$. Therefore, the state space of FRONTIER was also continuous.

As seen in the policy network diagrams, the feature maps from the CNN block were flattened along with the additional three feature vectors to produce the next fully connected layer consisting of $k(L-\tau+1)+3n+1$ neurons. This layer was followed by another fully connected layer consisting of $3(n+1)$ neurons. These fully connected layers had the rectified linear unit (ReLU) activation function and led to the final fully connected output layer as described earlier.

The hyperparameters of the policy network were selected somewhat arbitrarily with $L=20$ to represent 20 working days or one month and $\tau=5$ to represent a single working week. The amount of feature maps produced by the CNN block was chosen to be {$k=(n+1)$}. These and other policy network hyperparameters were not fine-tuned or optimised further in any way, partially due to time/computation constraints and partially to avoid over-fitting to any specific markets or asset portfolios.

\begin{figure*}[th]
\centerline{\includegraphics[width=0.65\textwidth]{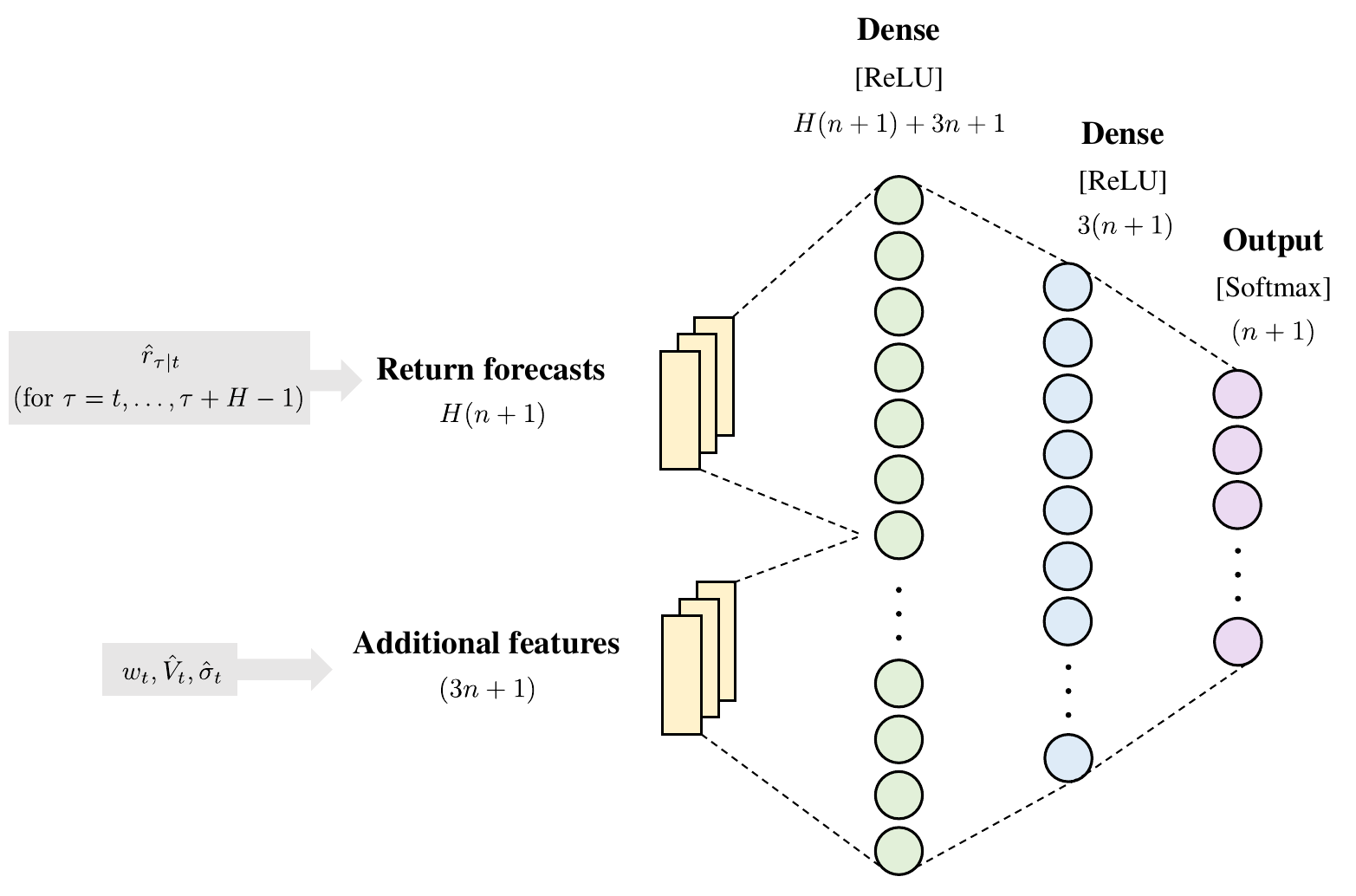}}
\caption{The forecast-only policy network of the FRONTIER model with only return forecasts and additional features as state inputs. The forecasts were explicit returns forecasts for all assets in the portfolio for $H$ steps into the future. The additional feature vectors gave the model the capacity to take transaction costs into account. The return forecasts and additional features were flattened and connected with fully-connected layers to produce the next portfolio weight vector as an output.}
\label{fig:policy-str-fcast}
\end{figure*}

The second policy network version was called the \textit{forecast-only} policy network (see diagram in Fig.~\ref{fig:policy-str-fcast}). This network had the same inputs as the log-returns policy network, with the log-returns window replaced by explicit returns forecasts of all assets in the portfolio. These forecasts took the form of $H$ vectors of $\hat{r}_{\tau|t} \in \mathbb{R}^{n+1}$, where each vector represented the returns forecast of a separate time-step. In this study, whenever return forecasts were given as state inputs to the FRONTIER model, a value of $H=2$ was used to ensure a valid comparison to the MPO could be made.

\begin{figure*}[th]
\centerline{\includegraphics[width=0.65\textwidth]{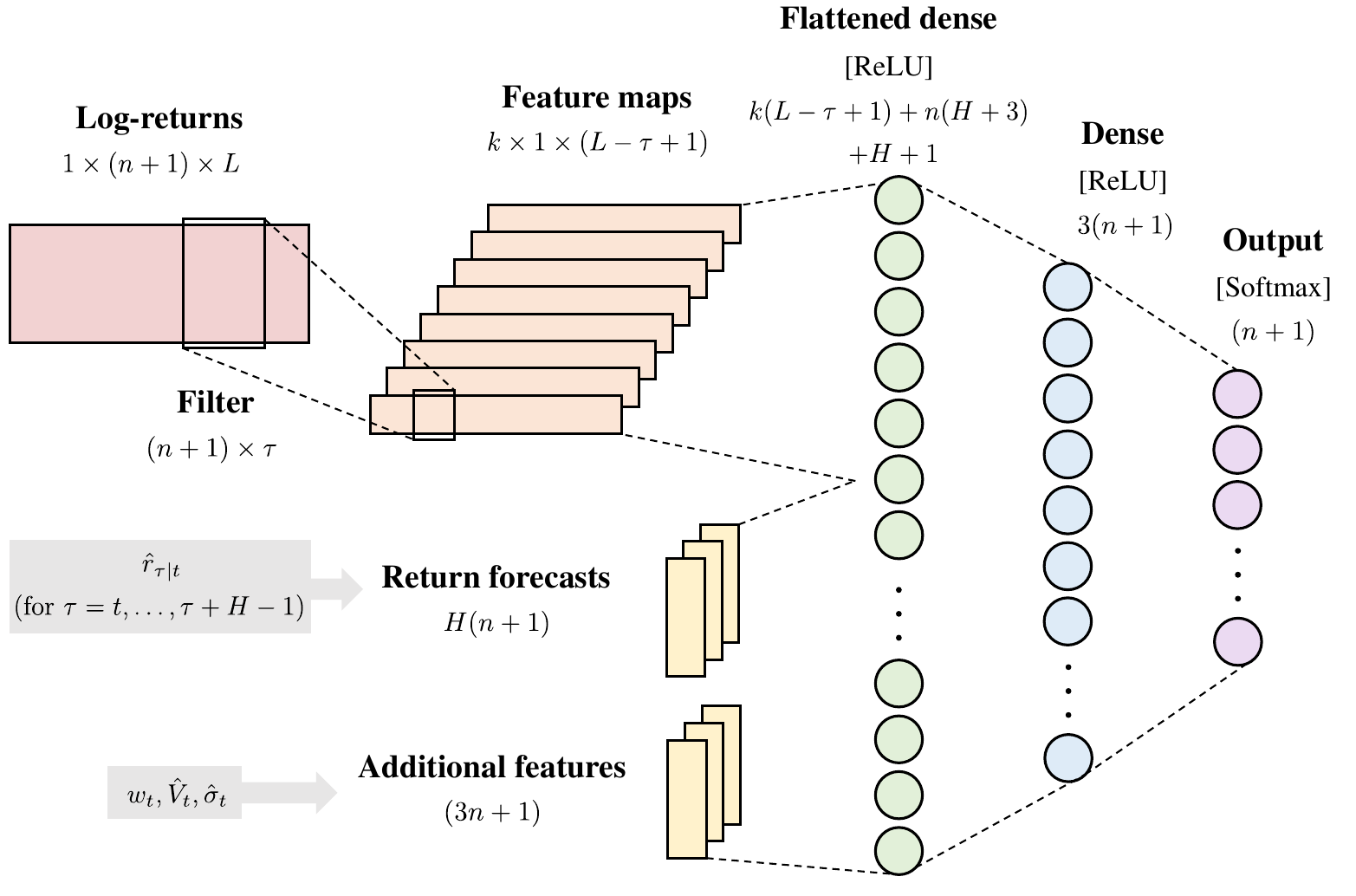}}
\caption{The all-inputs policy network of the FRONTIER model with all state inputs (historical log-returns, explicit return forecasts, and additional features). A convolutional filter was passed over the log-returns window to produce several feature maps. Explicit returns forecasts were also given for all assets in the portfolio for $H$ steps into the future. The additional feature vectors gave the model the capacity to take transaction costs into account. The feature maps, explicit return forecasts, and additional features were flattened and connected with fully-connected layers to produce the next portfolio weight vector as an output.}
\label{fig:policy-all-inp}
\end{figure*}

The forecast-only policy network was introduced to isolate the part of the policy network that produced forecasts. This way, the performance of the forecast-only policy could be compared to the log-returns policy to assess the efficacy of the CNN block in producing implicit returns forecasts. The third and final version, called the \textit{all-inputs} policy network (see diagram in Fig.~\ref{fig:policy-all-inp}), was a combination of the previous two policy networks such that it had access to all available state inputs. This version was introduced as a final check to see if any relative performance differences between the log-returns policy and the forecast-only policy were due to independent factors or if extra performance gains could be achieved by allowing access to all state input variables.

The FRONTIER model was trained in an episodic fashion where an episode of fixed length (30 days) was drawn from the training data according to a uniform distribution. The policy network parameters were then updated based on that episode's expected discounted future rewards. The discounted future rewards $G_t$ for each time step $t$ of the episode were calculated as follows~\cite{sutton-bartow}:
\begin{equation}
\label{eq:sum-disc-rew}
    G_t = \sum_{k=t+1}^{T} \gamma^{k-t-1} R_k
\end{equation}
where $\gamma=0.99$ is the future reward discount rate and $R_t$ is the immediate reward given at time $t$. This immediate reward was chosen to take the same form as the quantity maximised by the SPO and MPO mean-variance optimisation algorithms of Boyd \textit{et al.} (2017)~\cite{boyd} and is given in Equation~\eqref{eq:immediate-rew}. This expression was included and given the same form for two main reasons. Firstly, because no other existing RL methods take investor preferences into account and secondly, so that a valid comparison of FRONTIER could be made to SPO and MPO models for a range of investor preferences.
\begin{align}
\label{eq:immediate-rew}
R_t &= r^T_t w_{t+1} - \gamma^{\text{trade}}\phi^{\text{trade}}_{t} ( w_{t+1} -  w_{t}) \\
&\quad - \gamma^{\text{risk}} \psi_t (w_{t+1}) \nonumber
\end{align}

Finally, to get the expected discounted future rewards (from which the model parameters are updated during training), the average discounted future rewards were taken for each episode. 

\subsection{Data Collection}
All data used in this study came from Yahoo Finance~\cite{yahoofin} and Qunadl~\cite{quandl}. Data downloaded from Yahoo Finance included the daily open, high, low, and close prices of all assets, including their daily volume traded. Quandl was used to obtain the returns of what was considered the cash or risk-free asset in all portfolios. The US Federal three-month treasury bill rate was selected to be the risk-free asset.

The above-mentioned data were obtained for three different markets so that all models could be tested on different market conditions. The main goal was to select three markets: one where the overall market trend was upward; one where the overall trend was downward; and one where the overall trend was stable or sideways. A secondary goal was to select markets where these trend conditions were present for sufficiently long periods so that they could span both training and testing periods. A summary of the three selected markets is given in Table \ref{tab:market-data}. The price change and overall trends of these markets can also be seen in Fig.~\ref{fig:indices}.

\begin{table}[th]
\caption{Description of data for each market, including the overall trend, training period, testing period, and number of assets used.}
\centering
\resizebox{\columnwidth}{!}{%
\begin{tabular}{l|lllc} 
\hline
\textbf{Market} & \textbf{Trend} & \textbf{Training} & \textbf{Testing} &  \textbf{Assets}$^{\mathrm{a}}$ \\
\hline
\hline
Dow 30           &   Upward & \makecell[l]{2010-01-01 --\\2018-01-01} & \makecell[l]{2018-01-01 --\\2020-01-01} & 30 \\
Nikkei 225       & Sideways & \makecell[l]{2013-05-01 --\\2018-01-01} & \makecell[l]{2018-01-01 --\\2020-01-01} & 24 \\
Latin America 40 & Downward & \makecell[l]{2010-03-01 --\\2014-12-01} & \makecell[l]{2014-12-01 --\\2016-01-01} & 24 \\
\hline
\multicolumn{5}{l}{$^{\mathrm{a}}$Final number of assets selected after filtering and processing.}
\end{tabular}}
\label{tab:market-data}
\end{table}

\subsection{Data Processing}

For consistency and valid comparison, whenever models relied on estimates in this study, the same method was used in estimating values. This section describes these estimation methods and other data processing methods used in this study.

Whenever asset return forecasts were made, the method of Boyd \textit{et al.} (2017)~\cite{boyd} was followed. This forecasting method entailed perturbing realised future returns by adding noise with zero mean to simulate return forecasts. This forecasting method was employed in an attempt to keep the focus of this study on what is possible once return forecasts were already obtained. More specifically, the return forecasts were obtained for all non-cash assets using:
\begin{equation}
\label{eq:general-ret-fcast}
    \hat{r}_t = \alpha \left( r_t + \epsilon_t \right)
\end{equation}
where $\epsilon_t \sim \mathcal{N} \left( 0, \sigma^2_\epsilon \right)$ was zero-mean normally distributed noise and $\alpha$ was selected to minimise the mean squared error between the the realised returns $r_t$ and the noisy ``forecast'' $\hat{r}_t$, which equates to a scaling value of $\alpha = \sigma^2_r / (\sigma^2_r + \sigma^2_\epsilon)$, where $\sigma^2_r$ is the variance of $r_t$. A noise value of $\sigma^2_\epsilon=0.02$ was used along with a typical value of $\sigma^2_r=0.005$. This noise addition relates to a standard deviation in the forecast of 10 times that of the returns, which in turn relates to a return forecast accuracy on the higher end of what is expected in practice~\cite{boyd}.

In order to estimate the remaining values used in SPO and MPO for estimating the transaction cost and risk, the following steps were followed, again using the same method as Boyd \textit{et al.} (2017)~\cite{boyd}. Estimates for return volatility $\hat{\sigma}_t$ and volume traded $\hat{V}_t$ were calculated for each asset by taking a trailing 10-day moving average using the following equations:
\begin{align}
\label{eq:rolling-volume}
    \hat{V}_t &= \frac{1}{10} \sum^{10}_{\tau=1} V_{t-\tau} \\
\label{eq:rolling-volatility}
    \hat{\sigma}_t &= \frac{1}{10} \sum^{10}_{\tau=1} \sigma_{t-\tau}
\end{align}

For the additional features used as state inputs to the RL models, the same methods were used to calculate estimates for volume traded $\hat{V}_t$ and returns volatility $\hat{\sigma}_t$ as in Equation~\eqref{eq:rolling-volume} and Equation~\eqref{eq:rolling-volatility}, respectively. Before these additional state inputs were passed to the policy network, they were normalised to be on the same order of magnitude as $w_t$. This normalisation was done for each asset by dividing all $\hat{V}_t$ and $\hat{\sigma}_t$ values by their respective averages over the 30 days preceding the start of the training period specified in Table \ref{tab:market-data}.

In order to calculate the realised transaction costs, the realised volume traded $V_t$ was used along with the daily asset returns volatility $\sigma_t$. Since the collected data was on a daily frequency, $\sigma_t$ was approximated as in Boyd \textit{et al.} (2017)~\cite{boyd} using:
\begin{equation}
\label{eq:log-volatility}
    \sigma_t = \left| \log{\left( p_t^{\text{open}} \right)} - \log{\left( p_t^{\text{close}} \right)} \right|
\end{equation}
where $p_t^{\text{open}}$ and $p_t^{\text{close}}$ are the daily open and close prices of the asset in question, respectively.

\subsection{Experimental Testing and Comparison}
In multi-objective optimisation problems with several conflicting objectives, a set of viable solutions can be found where none of the solutions are dominated by others. These non-dominated solutions are all optimal solutions with trade-offs in at least one objective. Together, these non-dominated solutions form a multi-dimensional \textit{Pareto optimal frontier}~\cite{gg}. In this study, the experimentally derived set of optimal portfolios was referred to as the Pareto optimal frontier which is similar to the efficient frontier described by Markowitz~\cite{x,y}.

To compare the performance of all models against each other, they were all backtested on the testing portion of the data set for each market as specified in Table \ref{tab:market-data}. This test portion of the data set was kept from all models during the training phase to assess the out-of-sample performance of all models.  

The FRONTIER models were trained and tested, along with the SPO and MPO models for the entire investor preference spectrum spanned by the 504 combinations of risk-aversion $\gamma^{\text{risk}}$ and trade-aversion $\gamma^{\text{trade}}$ parameters. In addition to this, due to the stochastic nature of the RL model training process, all FRONTIER models were trained and tested on the same data set 10 times using different seed values for their pseudo-random number generating processes. This repetition was done for two reasons. Firstly, to assess the average performance of each model and secondly, to quantify the variance of the experimental performances obtained.

Each model's performance was assessed in terms of excess return $\overline{R^e}$ and excess risk $\sigma^e$ for a specific investor preference parameterised by $\gamma^{\text{risk}}$ and $\gamma^{\text{trade}}$. The excess risk and return was obtained using Equation~\eqref{ret-excess} and Equation~\eqref{sharpe}. This computation was done for all 504 parameter combinations expressing a range of investor preferences. Once all these points in risk-return space were obtained, the non-dominated set was determined by choosing all points for which the maximum excess return was obtained without increasing excess risk. This non-dominated constituted the Pareto optimal frontier, which was a set of optimal portfolios to hold during the test period.

Since the experiment was repeated 10 times for each FRONTIER model, the calculation of the Pareto frontier was done 10 times as well. The mean Pareto frontier was then calculated along with a t-test to determine the 95\% confidence interval of this mean Pareto frontier.

\subsection{Software, Libraries and Hardware}
All models used in this study, including traditional mean-variance optimisation models and RL models,  were programmed in Python 3. These models were trained and tested on a computer with 64 CPU cores and 252GB of RAM using the Ubuntu 21.04 (x86-64) operating system.

Our RL (FRONTIER) models were programmed in Python 3.7, using the Tensorflow 2.4 library~\cite{tensorflow-lib} for creating the different policy networks.

The SPO and MPO models were created in Python 3.6 and used the \textit{cvxpy} library~\cite{cvxpy-lib} for convex optimisation. These models were implemented using the \textit{cvxportfolio} library developed by Boyd \textit{et al.}~\cite{boyd-github}. In particular, an updated version, modified by Razvan Oprisor, was used~\cite{roprisor-github}. The EW model was also implemented using this updated \textit{cvxportfolio} library.

The state-of-the-art RL models (A2C, PPO, and DDPG) were implemented using the \textit{FinRL} library developed by Liu \textit{et al.}~\cite{finrl-lib}. Our implementation included minor modifications to simulate more realistic non-linear transaction costs. This library was created using Python 3.6 and used RL algorithms from the \textit{Stable-Baselines3} package~\cite{sb3-lib}.

\section{Results and Discussion}
\begin{figure*}[th]
\centerline{\includegraphics[width=0.95\textwidth]{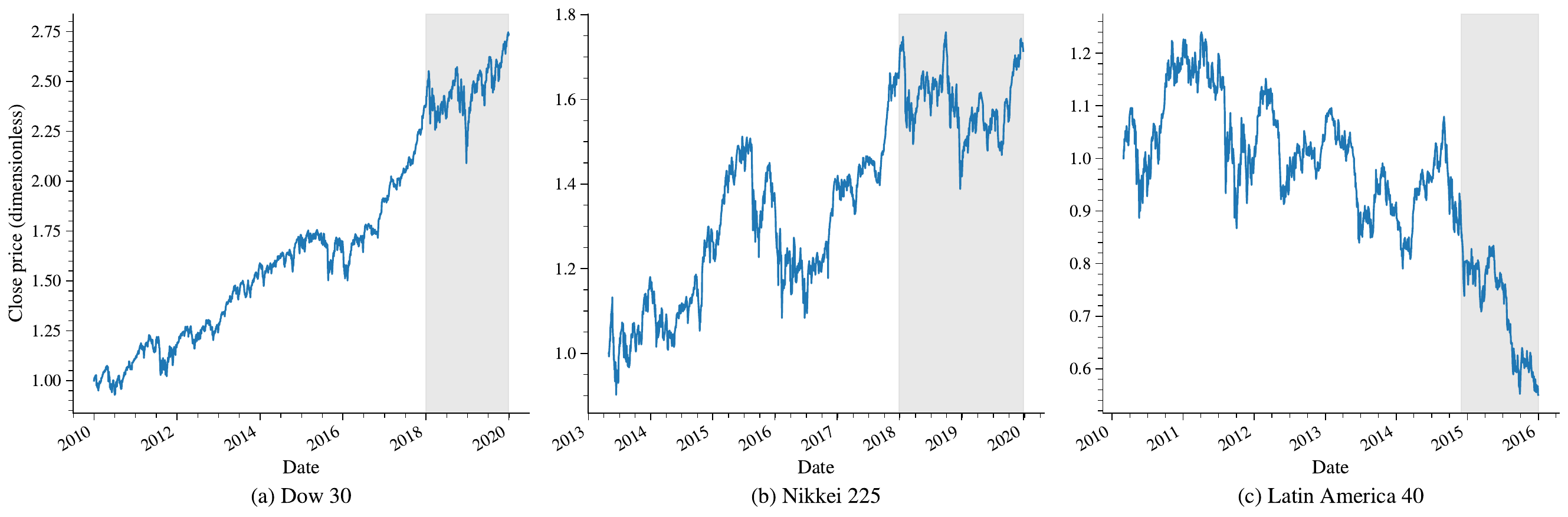}}
\caption{Non-dimensionalised price of all three markets used in this study. The shaded grey area indicates the test period of the data and the non-shaded white area indicates the training period of the data. The price changes are as follows for each period: Dow 30: +141.06\% (train) and +15.33\% (test); Nikkei 225: +58.48\% (train) and +0.49\% (test); Latin America 40: -25.36\% (train) and -43.11\% (test).}
\label{fig:indices}
\end{figure*}
To fully appreciate the results obtained by all the models in this study, it is helpful to consider the overall market trends during the training and testing periods of the data sets used. Fig.~\ref{fig:indices} shows the daily closing price of the market indices from which the stocks were selected.  Note that all prices were normalised and made dimensionless for easy comparison; this was done by dividing by the initial price. The shaded grey area on each chart indicates the testing period, whereas the non-shaded area indicates the training period. 

The upward trending market (Dow 30) shows a strong upward trend during the training and testing periods, with a price increase of just over 141\% during the training period and just over 15\% during the testing period. The sideways trending market (Nikkei 225) shows a slight upward trend during the training period with a price change of just over 58\% with a minor increase of under 1\% during the testing period. Finally, the downward trending market (Latin America 40) shows a strong downward trend with a change of just over 25\% during the training period, continuing downward with a price decrease of just over 43\% during the test period.

\begin{figure*}[th]
\centerline{\includegraphics[width=0.95\textwidth]{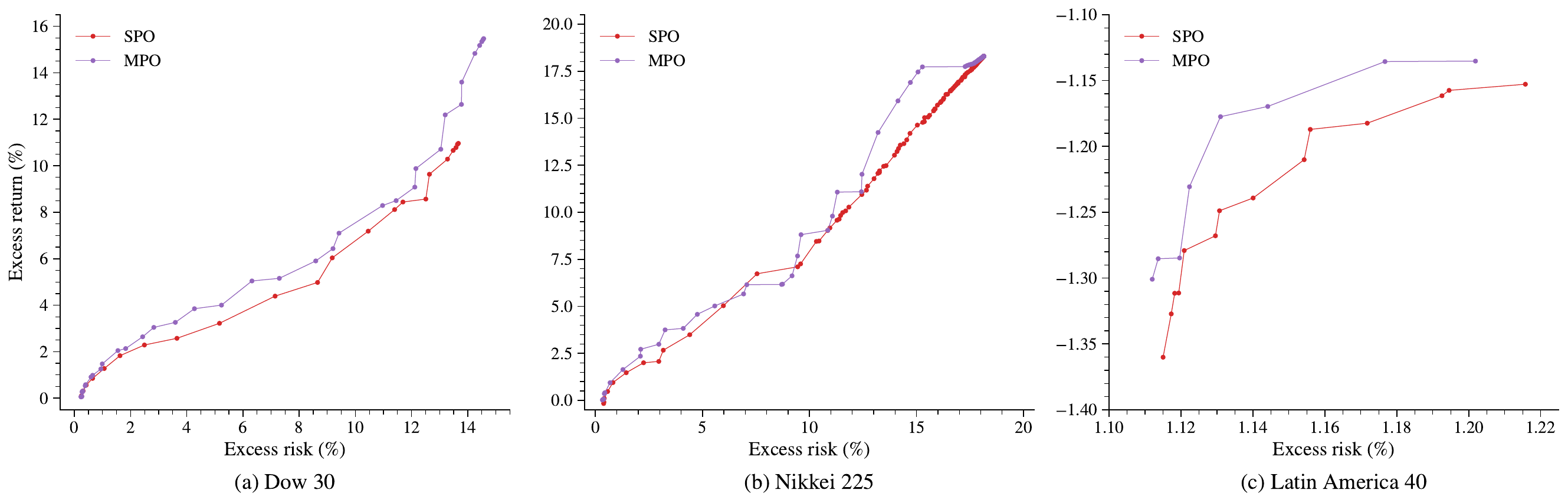}}
\caption{Pareto optimal frontiers in risk-return space of SPO and MPO models on all three markets produced by parameter sweep of all 504 pairwise combinations of risk and trade-aversion. MPO outperforms SPO on average in all three markets. Note: significant scale difference on (c) Latin America 40.}
\label{fig:boyd-only}
\end{figure*}
Fig.~\ref{fig:boyd-only} shows the performance of SPO and MPO on each of the three markets in isolation. These are the Pareto optimal frontiers obtained by simulating backtests over the test period for all 504 pairwise combinations of risk-aversion $\gamma^{\text{risk}}$ and trade-aversion $\gamma^{\text{trade}}$. This figure shows, as expected that MPO slightly outperforms SPO on average in all three markets. This outperformance is likely due to the extra time-step taken into account by the MPO model during its multi-period optimisation. This result is also found in the study of Boyd \textit{et al.}(2017) on the S\&P500 market. 

It is important to note the difference in scale in Fig.~\ref{fig:boyd-only}(c) of the plot for the Latin America 40 market. Both SPO and MPO produced only negative excess returns over a very small excess risk range. Upon closer inspection of the portfolio weight vectors $w_t$ these models produced, it is clear that both SPO and MPO almost exclusively invested in the risk-free asset, only to shift to small positions in riskier stocks for very short periods as their risk-aversion decreased. This shift explains the slight variation of excess risk and return that produced these Pareto frontiers. However, both SPO and MPO behaved differently in the the Dow 30 and Nikkei 225 markets. In these two markets, a more gradual portfolio weight change was made, spanning the whole spectrum from fully invested in the risk-free asset to large positions in risky stocks as the risk-aversion parameter was lowered. Larger trades (changes in $w_t$) were also observed as the trade-aversion parameter was reduced.

\begin{figure*}[th]
\centerline{\includegraphics[width=0.95\textwidth]{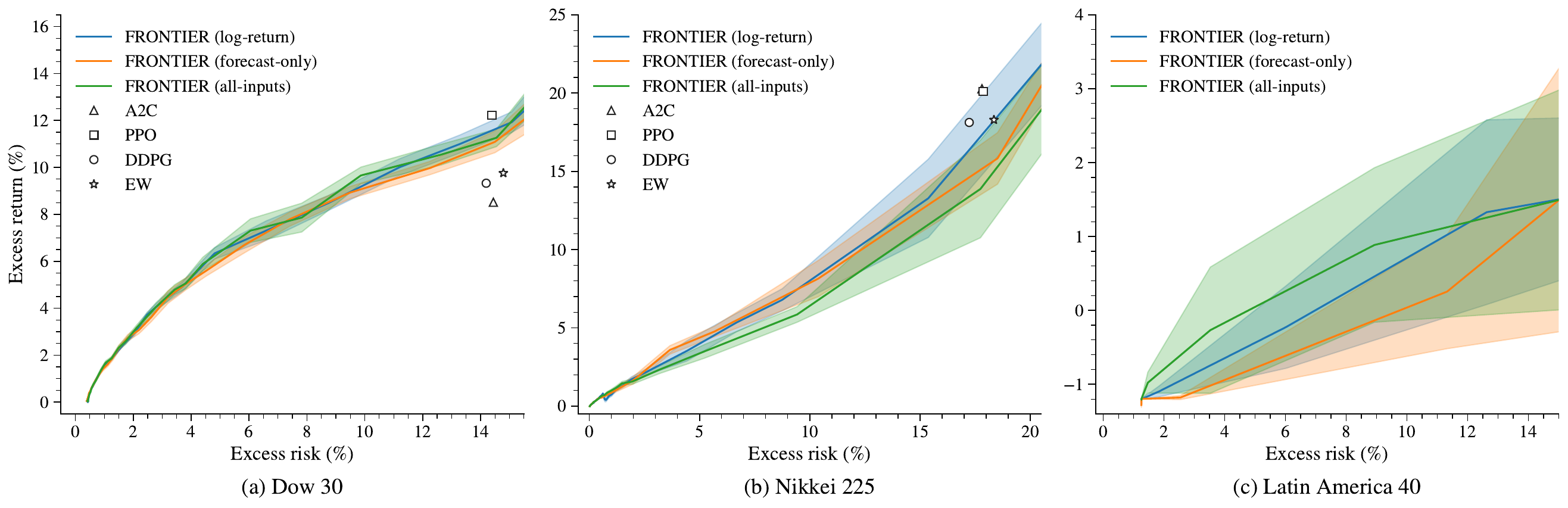}}
\caption{Pareto optimal frontiers (mean with 95\% confidence interval) in risk-return space of the FRONTIER model with three different policy networks on all three markets. Frontiers were produced by parameter sweep of all 504 pairwise combinations of risk and trade-aversion. Also included are the performances of three state-of-the-art RL models (A2C, PPO, and DDPG) along with the EW portfolio. Note: A2C, PPO, DDPG, and EW are not shown for (c) Latin America 40 since they produced large negative returns and was off the chart area.}
\label{fig:rl-only}
\end{figure*}
\begin{figure*}[th]
\centerline{\includegraphics[width=0.95\textwidth]{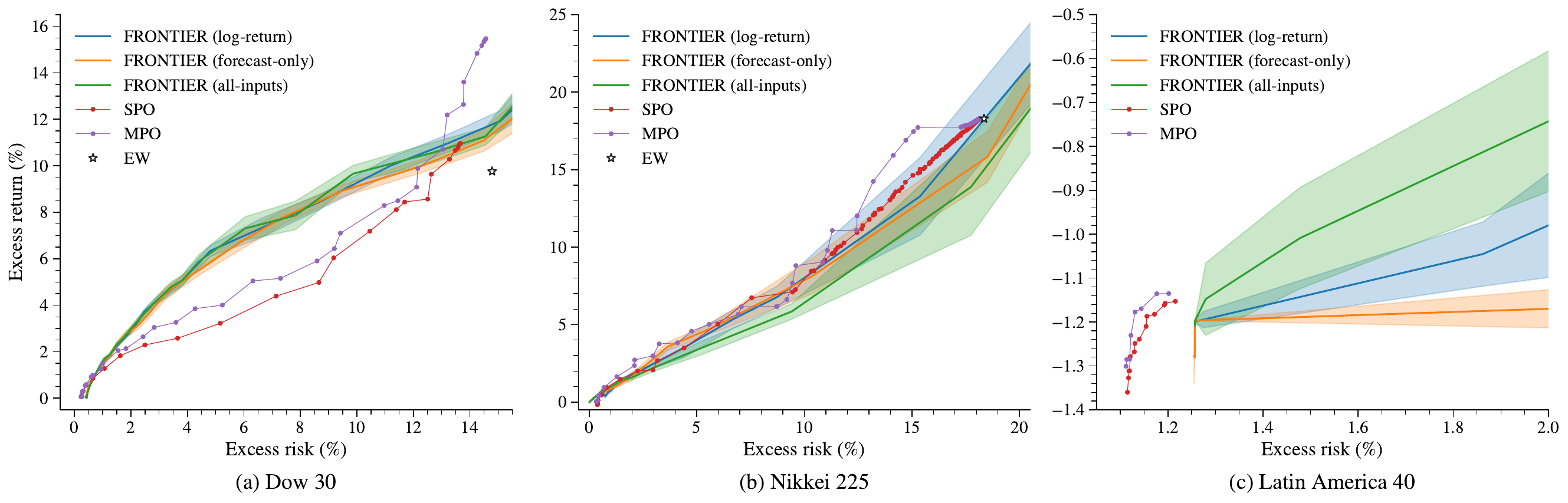}}
\caption{Direct comparison of Pareto optimal frontiers in risk-return space of FRONTIER models (mean with 95\% confidence interval) and convex mean-variance optimisation models (SPO and MPO). All frontiers were produced by the same parameter sweep of all 504 pairwise combinations of risk and trade-aversion. Also included is the EW portfolio performance for reference.}
\label{fig:boyd-vs-rl}
\end{figure*}

Fig.~\ref{fig:rl-only} shows the mean Pareto frontiers (along with their 95\% confidence intervals) produced by FRONTIER when using different policy network architectures. For the Dow 30 market, all three policy networks performed very similarly for the entire excess risk and return ranges. On the Nikkei 225 market, the performance of all three policy networks was also similar, with the mean frontiers of the log-return and forecast-only networks slightly outperforming the all-inputs network for the most part and the log-return network achieving the most excess returns towards the high-risk end. On the Latin America 40 market, all three policy networks were also very closely matched with the all-inputs version producing the highest excess returns towards the low-risk end (see Fig.~\ref{fig:boyd-vs-rl} for closer inspection on the low-risk end).

However, considering the overlapping confidence intervals for the vast majority of the frontiers in all three markets, it seems that none of these policy networks could significantly outperform any of the others consistently. This result suggests that the all-inputs policy network did not have an added advantage even with all state inputs at its disposal. It also suggests that the log-returns policy network implicitly produced asset return forecasts with the same degree of accuracy as the perturbed realised return forecasts.

Looking at the state-of-the-art RL models' performance in Fig.~\ref{fig:rl-only}, the same qualitative performance of the study by Yang \textit{et al.} (2020) is also found in the Dow 30 market in that PPO and DDPG both outperformed A2C for upward trending market conditions. In the Nikkei 225 market, these three models performed more similarly, with PPO and A2C having almost the same performance and DDPG producing slightly less excess returns for a similar risk value. These RL methods do not appear in the Latin America 40 market plot due to their large negative excess returns that are off the chart area (-28.4\% for DDPG; -29.4\% for PPO; and -35.5\% for A2C).

Fig.~\ref{fig:rl-only} also shows the performance of FRONTIER relative to A2C, PPO, and DDPG. In the Dow 30 market, FRONTIER could outperform both A2C and DDPG, with PPO producing slightly more returns than the upper confidence interval of FRONTIER fitted with a log-returns policy network. For the Nikkei 225 market, there is no significant performance difference between our RL equipped with a log-returns policy network and A2C, PPO, or DDPG. This result suggests that in markets with an upward trend, FRONTIER outperformed or at least closely matched the performance of state-of-the-art RL models seeking high returns. This result also suggests that in sideways trending markets, FRONTIER (with a log-returns policy network) matched the performance of state-of-the-art RL models seeking high returns.

In order to assess the effect that our non-linear transaction cost modification had on portfolio management performance, the DDPG, PPO, and A2C models from Yang \textit{et al.} (2020)~\cite{p} were evaluated using the different transaction cost functions. These models were selected because their original versions used linear transaction cost functions. For this comparison, all three models were trained and tested on the Dow 30 market for the same periods seen in Table~\ref{tab:market-data}. The original non-linear transaction cost function for these models was equivalent to using Equation~\ref{eq:transaction-cost} with values of $a=0.0005$, $b=0$, and $c=0$. These original versions were compared to our modified versions (seen in Fig.~\ref{fig:rl-only}(a)) with non-linear transaction cost functions ($a=0.0005$, $b=1$, and $c=0$). The excess returns, excess risk, and Sharpe ratio produced by these models can be seen in Table~\ref{tab:trans-cost}. For all three models, the excess risk achieved was similar when using the two different transaction cost functions (0.5\% average difference). However, there was a slight difference in the excess returns (1.4\% on average). PPO managed to produce slightly more excess returns using the non-linear transaction cost function, whereas DDPG and A2C both produced higher excess returns with the linear transaction cost function. PPO also achieved a slightly higher Sharpe ratio with the non-linear transaction cost function whereas DDPG and A2C produced higher values with the linear transaction cost function. This result suggests that the linear transaction cost function might overestimate risk-adjusted returns (Sharpe ratio) for some models like DDPG and A2C while slightly underestimating them for other models like PPO.

\begin{table}[th]
\caption{Change in excess returns, excess risk, and Sharpe ratio obtained by DDPG, PPO, and A2C model on the Dow 30 market when using linear and non-linear transaction cost functions.}
\centering
\resizebox{\columnwidth}{!}{
\begin{tabular}{l|lccc}
\hline
\makecell[l]{\textbf{Model}} &  \makecell[l]{\textbf{Transaction} \\\textbf{cost}}  &  \makecell[l]{\textbf{Excess} \\\textbf{return (\%)}} &  \makecell[l]{\textbf{Excess} \\\textbf{risk (\%)}} &  \makecell[l]{\textbf{Sharpe} \\\textbf{ratio}} \\
\hline
\hline
\multirow{2}{*}{DDPG} & Linear &             10.801 &           14.908 &         0.724 \\
    & Non-linear &              9.328 &           14.194 &         0.657 \\
\multirow{2}{*}{PPO} & Linear &             11.733 &           14.996 &         0.782 \\
    & Non-linear &             12.227 &           14.395 &         0.849 \\
\multirow{2}{*}{A2C} & Linear &             10.819 &           14.600 &         0.741 \\
    & Non-linear &              8.516 &           14.442 &         0.590 \\
\hline
\end{tabular}}
\label{tab:trans-cost}
\end{table}

Thus far, our results address the four limitations identified in the evaluated previous research looking at portfolio management using RL methods. These results provide insight into the performance of RL methods when a wide variety of investor preferences are considered in terms of risk and trade-aversion. These results produce an entire Pareto optimal frontier from which investors can choose their risk and trade-aversion parameters to suit their particular risk and return objectives. Moreover, these model performances are more realistic compared to the aforementioned prior research from a transaction cost perspective. This improvement comes from the inclusion of non-linear changes in the transaction cost introduced by market volatility and trading volume which was taken into account in addition to the linear changes related to bid-ask spreads and broker commission. Finally, the limitation of only testing on a single market was also addressed by conducting tests on three markets from different economies with different overall price trends. With these limits addressed, a more comprehensive comparison of traditional mean-variance optimisation methods could be made with RL methods and is considered next.

The performance of FRONTIER models is directly compared to that of the traditional mean-variance optimisation methods in Fig.~\ref{fig:boyd-vs-rl} for all three markets. In the Dow 30 market, FRONTIER could significantly outperform both SPO and MPO for excess risk values between around 1\% and 13\%. When taking on excess risk above around 13\%, however, MPO produced significantly higher returns. In the Nikkei 225 market, all FRONTIER models produced similar excess returns for the majority of the risk range. In this market, MPO produced significantly higher excess returns compared to FRONTIER models for excess risk values between around 13\% and 16\%. A direct comparison could not be made in the Latin America 40 market since there was no overlap in the FRONTIER models' Pareto frontiers and those of SPO or MPO. It might be possible to extend the Pareto frontiers of the SPO and MPO models to produce an overlapping area by testing a wider range of risk and trade-aversion parameters. In the parameter sweep tested, lower risk-aversion parameters did lead to points further to the right in this risk-return space. However, they all produced very low (and often negative) returns and were not Pareto optimal. These results suggest that FRONTIER is able to significantly outperform traditional mean-variance optimisation methods like SPO and MPO in upward trending markets up to some excess risk limit (in the case of the Dow 30 market, this limit was around 13\%). Our results also suggest that in sideways trending markets, the  performance of SPO and MPO can be closely matched by FRONTIER for the majority of the excess risk range tested. No conclusions could be drawn on the outperformance of traditional mean-variance optimisation models and FRONTIER in downward trending markets.

Given that the FRONTIER model with forecast-only policy network had the same state inputs as SPO and MPO, the main difference in these models were the temporal aspects of their optimisation algorithms. FRONTIER optimised its reward signal for expected future rewards over a period of 30 days (one episode length), where SPO and MPO only optimised their rewards over a period of one or two days. These results suggest that there is some advantage in using RL methods for portfolio management because of the way they optimise for expected future rewards over more extended periods of time (at least under certain market conditions).

Fig.~\ref{fig:boyd-vs-rl} shows that in the Nikkei 225 market, SPO, MPO, and FRONTIER models produced very similar returns to the EW strategy around the 18\% excess risk mark. Indeed, after inspecting the portfolio weight vectors of both SPO and MPO models, they seem to use a strategy very close to that of EW. In the case of FRONTIER models, this seems to be more of a coincidence as they were still predominantly invested in one to three risky assets and cash. For the Dow 30 market, all FRONTIER models and MPO could produce greater returns for a given amount of risk compared to the EW portfolio. Finally, in the Latin America 40 market, even though SPO, MPO, and FRONTIER produced mostly negative excess returns, they did learn to invest almost solely in the risk-free asset for high risk-aversion values. Therefore, SPO, MPO, and FRONTIER arguably outperform the EW strategy, which produced extreme negative excess returns (-29.9\%). These results suggest that in upward or downward trending markets, the EW strategy can be outperformed using SPO, MPO, or FRONTIER in terms of returns. It also suggests that it is not possible to significantly outperform the EW strategy in a sideways trending market using either traditional mean-variance optimisation or RL models from this study.

\section{Conclusion}

This study compared the portfolio management performance of traditional mean-variance optimisation models like SPO and MPO to that of RL methods (FRONTIER) in risk-return space. One of the main reasons for doing so was the capacity of RL models to optimise their expected rewards over more extended periods compared to the relative short-sighted optimisations of SPO and MPO. This long-term optimisation is important when considering the portfolio management problem since immediate actions can affect an agent's ability to produce optimal rewards in the future due to transaction costs. Before doing so, four limitations of the evaluated previous research on portfolio management with RL methods were addressed. These limitations were addressed in order to achieve a more comprehensive comparison. This process entailed creating our RL models that could take a wide range of investor preferences into account in terms of trade-aversion and risk-aversion to suit their particular risk and return objectives. The inclusion of these investor preference parameters into our RL models resulted in Pareto optimal frontiers in risk-return space that could be compared to those of traditional mean-variance optimisation models (SPO and MPO). Our tests were repeated on three different markets that represented three different economies and overall market trends to assess the applicability of our results to different market conditions. All models in this study were created/modified to account for more realistic non-linear changes in transaction cost introduced by market volatility and trading volume in addition to linear changes related to bid-ask spreads and broker commission.

The results of this study suggest that there can be an advantage to using RL methods compared to traditional mean-variance optimisation methods for portfolio management because they optimise for expected future rewards over more extended periods (at least under certain market conditions). Our RL models were able to significantly outperform traditional mean-variance optimisation methods like SPO and MPO in upward trending markets up to some excess risk limit (in the case of the Dow 30 market, this limit was around 13\%). Our results also suggest that in sideways trending markets, the  performance of SPO and MPO can be closely matched by our RL models for the majority of the excess risk range tested. In downward trending markets, no conclusions could be drawn on the outperformance of traditional mean-variance optimisation models and our RL models. The most benefit can be gained from our RL methods in upward trending markets as this is where they have the best potential to outperform EW and traditional mean-variance optimisation methods. This result especially applies to a particular excess risk range (in the Dow 30 market, this was between around 1\% and 13\%). This range might change depending on the market or underlying assets held in the portfolio.

The caveats and specific market conditions under which these models can outperform each other highlight the importance of a more comprehensive comparison in risk-return space for a range of risk values. These Pareto optimal frontiers give investors a more granular view of which models might provide better performance for their specific risk tolerance or return targets. It also gives insight to model developers to see where the possible limitations of specific methods are so that they can be improved.

In future work, the methods of this study can be repeated on different markets with similar overall price trends to see if our results hold for all markets with these characteristics. Our RL models can also be extended to allow short positions to see how this affects the results. It would also be interesting to add more features like market sentiment to the state input of RL models to see whether this improves the implicit returns forecasting ability and subsequent portfolio management performance. Perhaps the most interesting development from this study would be to change the reward function of other/future state-of-the-art RL models to incorporate specific investor preferences so that they can also be compared more comprehensively in risk-return space to traditional mean-variance optimisation methods.

\printbibliography[heading=bibintoc, title=REFERENCES]


\begin{IEEEbiography}[{\includegraphics[width=1in,height=1.25in,clip,keepaspectratio]{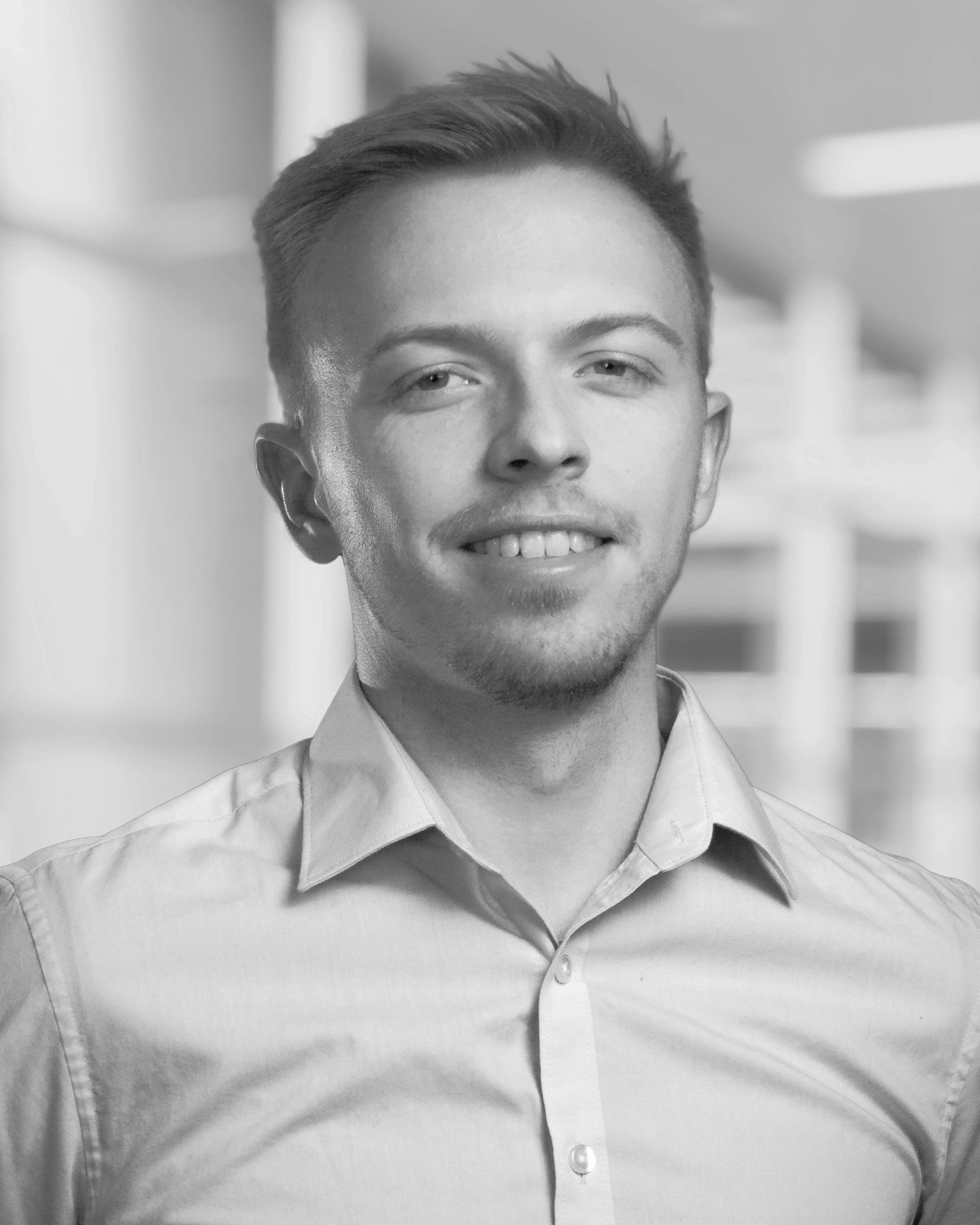}}]{Ruan Pretorius} received the BSc. Engineering degree in
mechanical engineering from the University of The Witwatersrand, South Africa, in 2019. He is currently
pursuing the MSc. degree in e-Science with the School of Computer Science and Applied Mathematics at the University of The Witwatersrand, South Africa. His research interests include artificial intelligence and machine learning methods, especially reinforcement learning, applied to quantitative finance and investment portfolio management.
\end{IEEEbiography}

\begin{IEEEbiography}[{\includegraphics[width=1in,height=1.25in,clip,keepaspectratio]{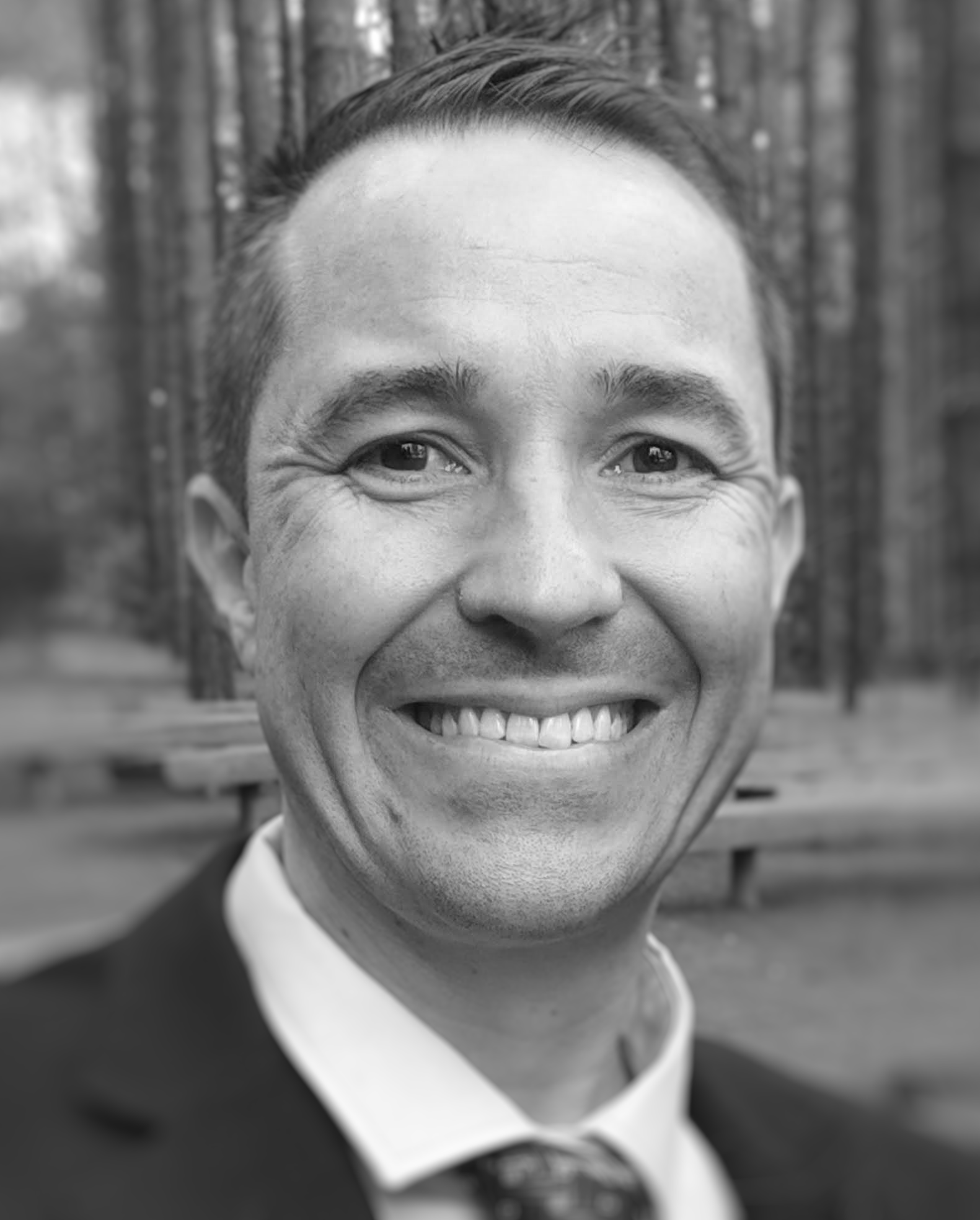}}]{Terence L. van Zyl}  (Member, IEEE) received the M.Sc. and Ph.D. degrees in computer science for his thesis on agent-based complex adaptive systems from the University of Johannesburg, Johannesburg, South Africa. He holds the Nedbank Research and Innovation Chair with the University of Johannesburg, where he is currently a Professor with the Institute for Intelligent Systems. He is also an NRF-rated scientist and he has more than 15 years of experience researching and innovating large scale streaming analytics systems for government and industry. His research interests include data-driven science and engineering, prescriptive analytics, machine learning, meta-heuristic optimisation, complex adaptive systems, high-performance computing, and artiﬁcial intelligence.
\end{IEEEbiography}

\vfill

\end{document}